\documentclass[conference]{IEEETran}

\usepackage{url}
\usepackage{svg}
\usepackage{acronym}
\usepackage{subcaption}
\usepackage{cite}
\usepackage[export]{adjustbox}
\usepackage{amsmath,amssymb,amsfonts}
\usepackage{graphbox}
\usepackage{multirow}
\usepackage{textcomp}
\usepackage{listings}
\usepackage{enumitem}
\usepackage{courier} %% Sets font for listing as Courier.
\usepackage{xcolor}
\usepackage{caption}
\usepackage{booktabs} % Required for better horizontal rules in tables
\usepackage{fouriernc} % Use the New Century Schoolbook font
\usepackage{graphicx,wrapfig,lipsum}
\usepackage[colorinlistoftodos]{todonotes}
\usepackage[colorlinks]{hyperref}
\usepackage[linesnumbered,ruled]{algorithm2e}

\begin{document}

\title{The Importance of Worst-Case Memory Contention Analysis for Heterogeneous SoCs}

\author{Lorenzo Carletti, Gianluca Brilli, Alessandro Capotondi, Paolo Valente, Andrea Marongiu}

\maketitle

\begin{abstract}
Memory interference may heavily inflate task execution times in Heterogeneous Systems-on-Chips (HeSoCs). Knowing worst-case interference is consequently fundamental for supporting the correct execution of time-sensitive applications. In most of the literature, worst-case interference is assumed to be generated by, and therefore is estimated through read-intensive synthetic workloads with no caching.

Yet these workloads do not \emph{always} generate worst-case interference. This is the consequence of the general results reported in this work. By testing on multiple architectures, we determined that the highest interference generation traffic pattern is actually hardware dependant, and that making assumptions could lead to a severe underestimation of the worst-case (in our case, of more than 9$\times$).
\end{abstract}
\section{Introduction}
\label{sec:intro}

Heterogeneous on-chip Systems (HeSoC) combine the benefits of low power consumption Systems on Chip (SoC) with the ability to use accelerators to execute specialized workloads efficiently. Commercial-off-the-shelf (COTS) HeSoCs often make use of GP-GPU or FPGA as accelerators, combined with multi-core general-purpose \textit{host} CPUs. This provides both specialization and flexibility.

\begin{figure}
    \centering
    \includegraphics[width=0.5\textwidth]{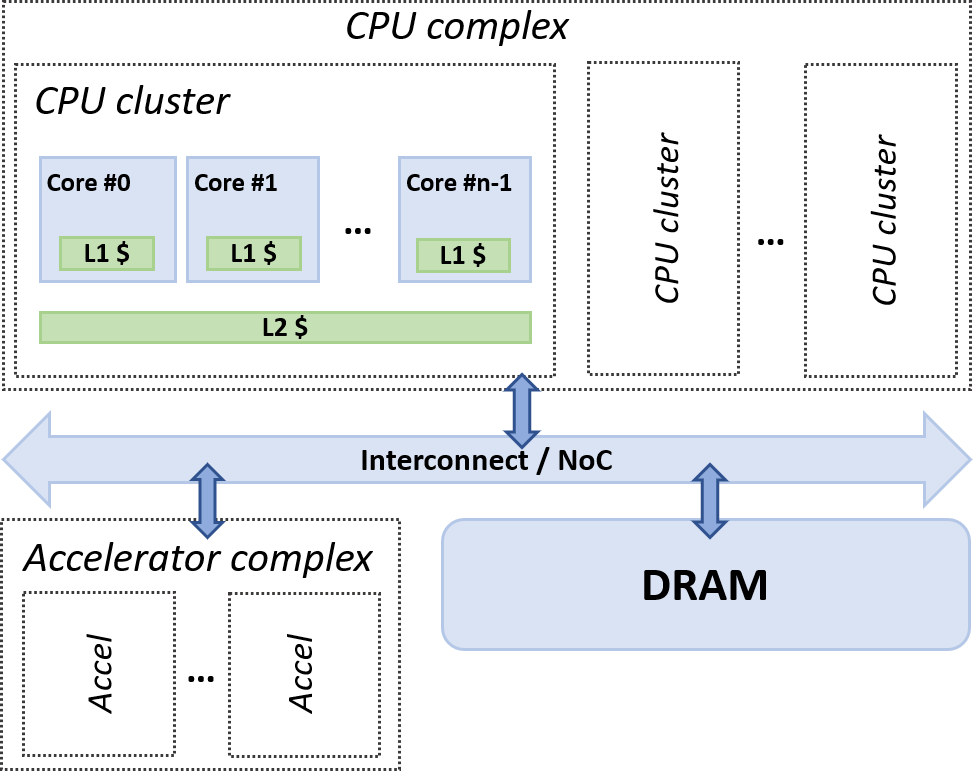}
        \caption{Shared memory architectural template.}\label{fig:hesoc}
 \end{figure}

These systems typically rely on a shared-memory organization, where the aforementioned compute units are interconnected through a shared bus to the main system DRAM, as shown in Fig. \ref{fig:hesoc}. 
As the number of compute engines grows in the chip, the main memory is subject to increasing contention.
 
This potentially causes the tasks executing on the various units to experience decreased bandwidth and, as a consequence, an increased execution time due to mutual interference~\cite{9774768}. This is particularly problematic for time-sensitive applications.
The problem has been extensively studied before~\cite{memory_subsystem, unexpected_diversity, read_bcs, 6214768, 9203722, 9523780}, and several different approaches to mitigate the effects of memory interference have been developed ~\cite{prem,memguard,injection, eth_interf,10.1145/3548773}.

In order to prove the validity of an interference mitigation solution, it's important to have a proper understanding of the timing effects (i.e., slowdown) a workload under test experiences in the worst case, and showing that a particular approach can handle such a scenario. We define as worst-case the program which is slowed down the most by other tasks causing DRAM interference.

\begin{figure*}[!htb]
    \begin{subfigure}[c]{.3\textwidth}
        \centering
        % include first image
        \caption{\textbf{Interf=READ\_MISS}}
        \includegraphics[width=\columnwidth]{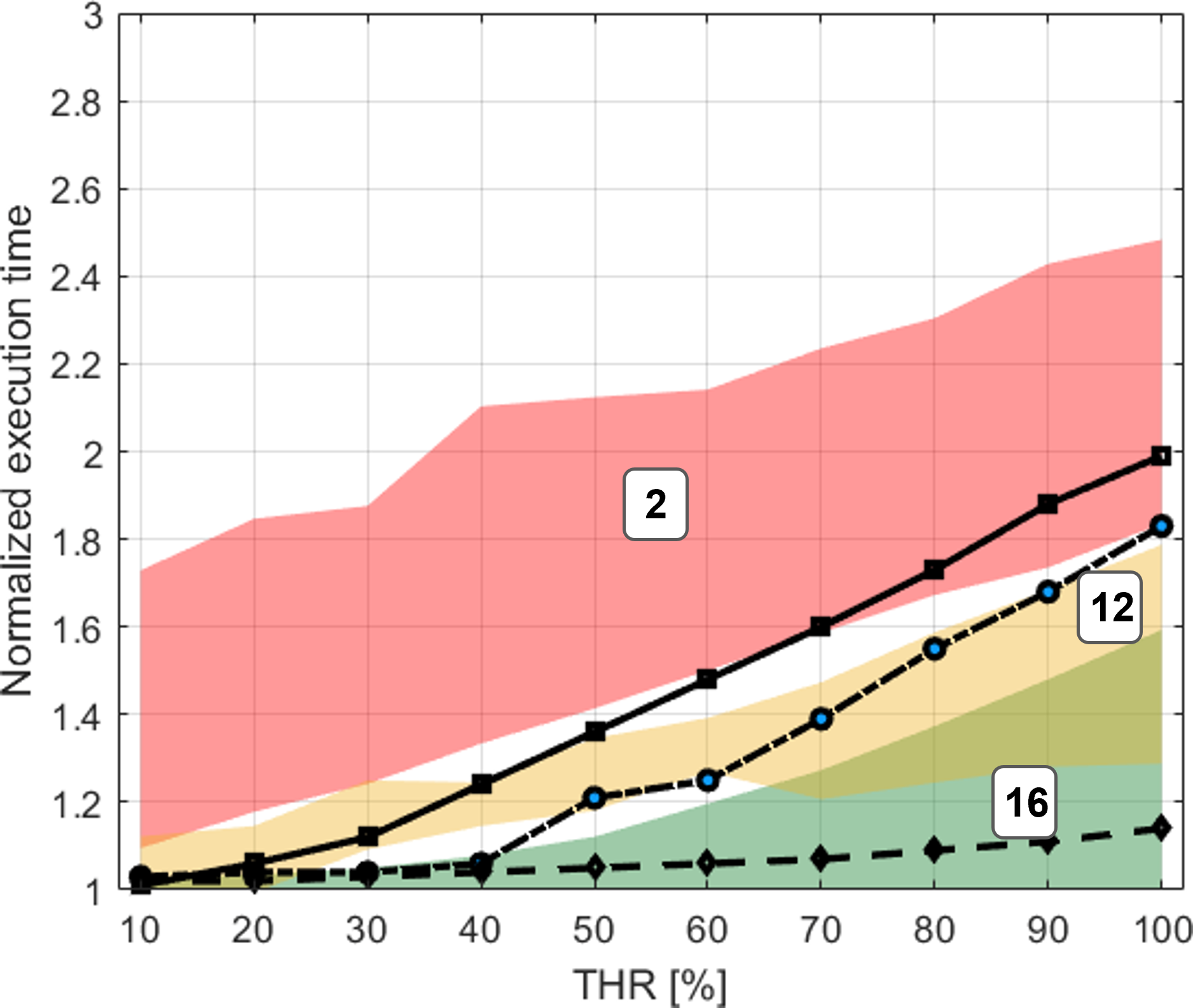}  
        \label{fig:NVIDIAa}
    \end{subfigure}
    \begin{subfigure}[c]{.28\textwidth}
        \centering
        % include first image
        \caption{\textbf{Interf=MEMSET}}
        \includegraphics[trim=20 0 0 0,clip,width=\columnwidth]{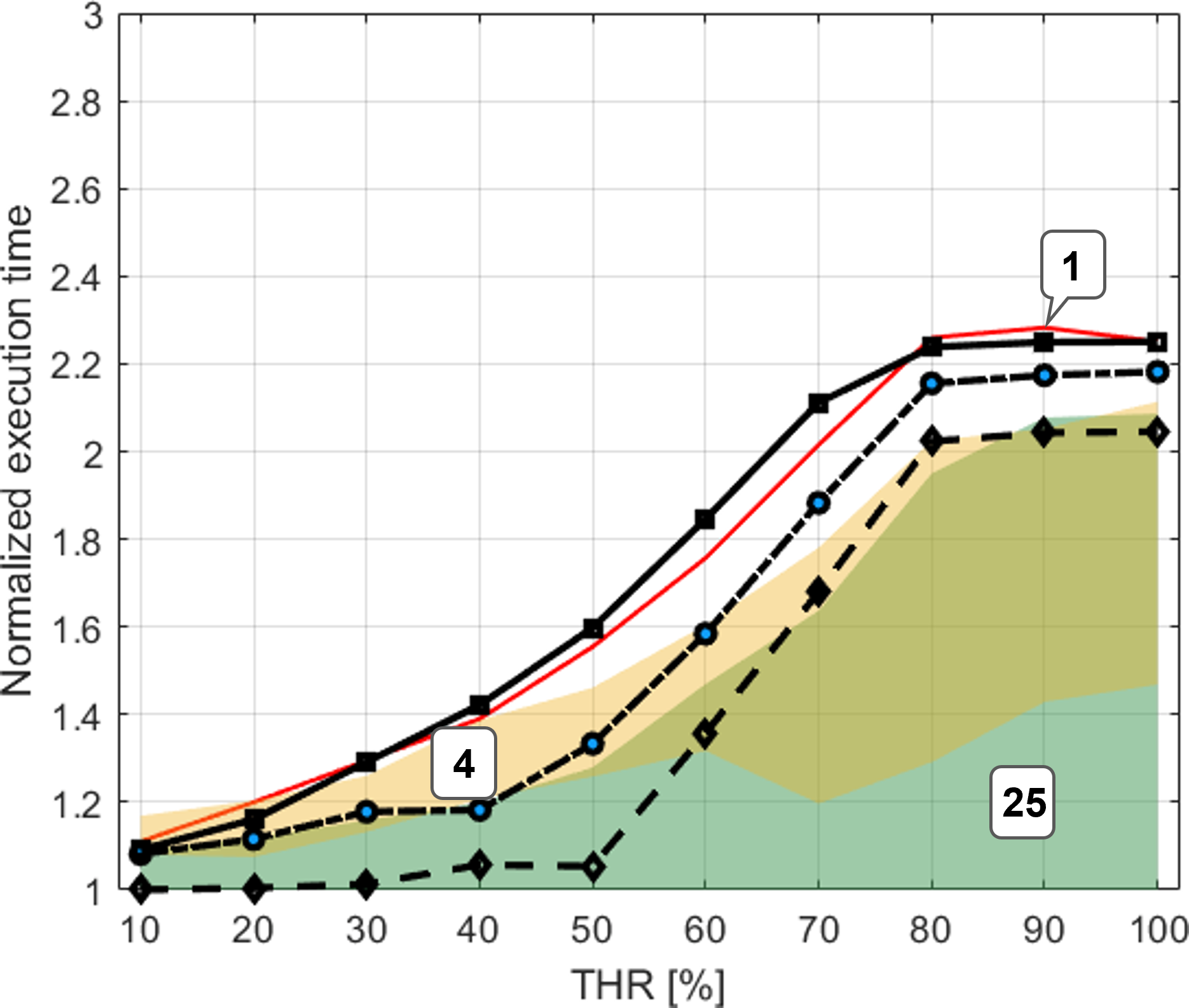}  
        \label{fig:NVIDIAb}
    \end{subfigure}
    \begin{subfigure}[c]{.28\textwidth}
        \centering
        % include first image
        \caption{\textbf{Interf=MEMCPY}}
        \includegraphics[trim=20 0 0 0,clip,width=\columnwidth]{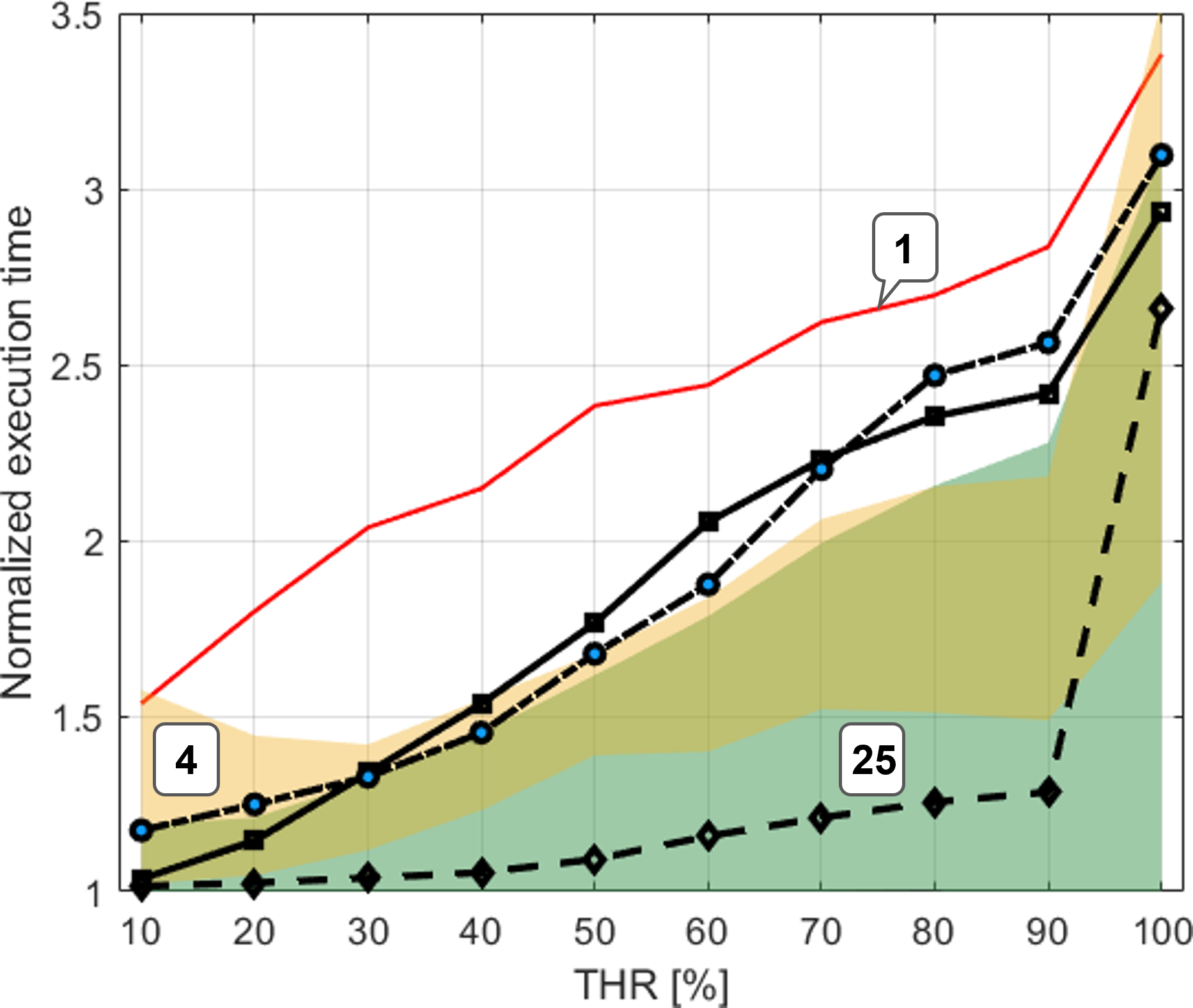}
        \label{fig:NVIDIAc}
    \end{subfigure}
    \begin{subfigure}[c]{.1\textwidth}
        \centering
        % include third image
        \includegraphics[width=\columnwidth]{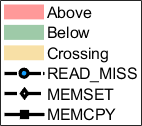}
        \label{fig:_tmp8}
    \end{subfigure}
    \caption{\textbf{NVIDIA TX2}. Execution time increase of the three synthetic benchmarks (curves) and the Polybench benchmarks (solored areas) running on the core under test with increasing interference from the other cores (THR\%). The workload executed by the interference cores is indicated above the plot.}
    \label{fig:interf_NVIDIA}
\end{figure*}

\begin{figure*}[!htb]
    \begin{subfigure}[c]{.3\textwidth}
        \centering
        % include first image
        \caption{\textbf{Interf=READ\_MISS}}
        \includegraphics[width=\columnwidth]{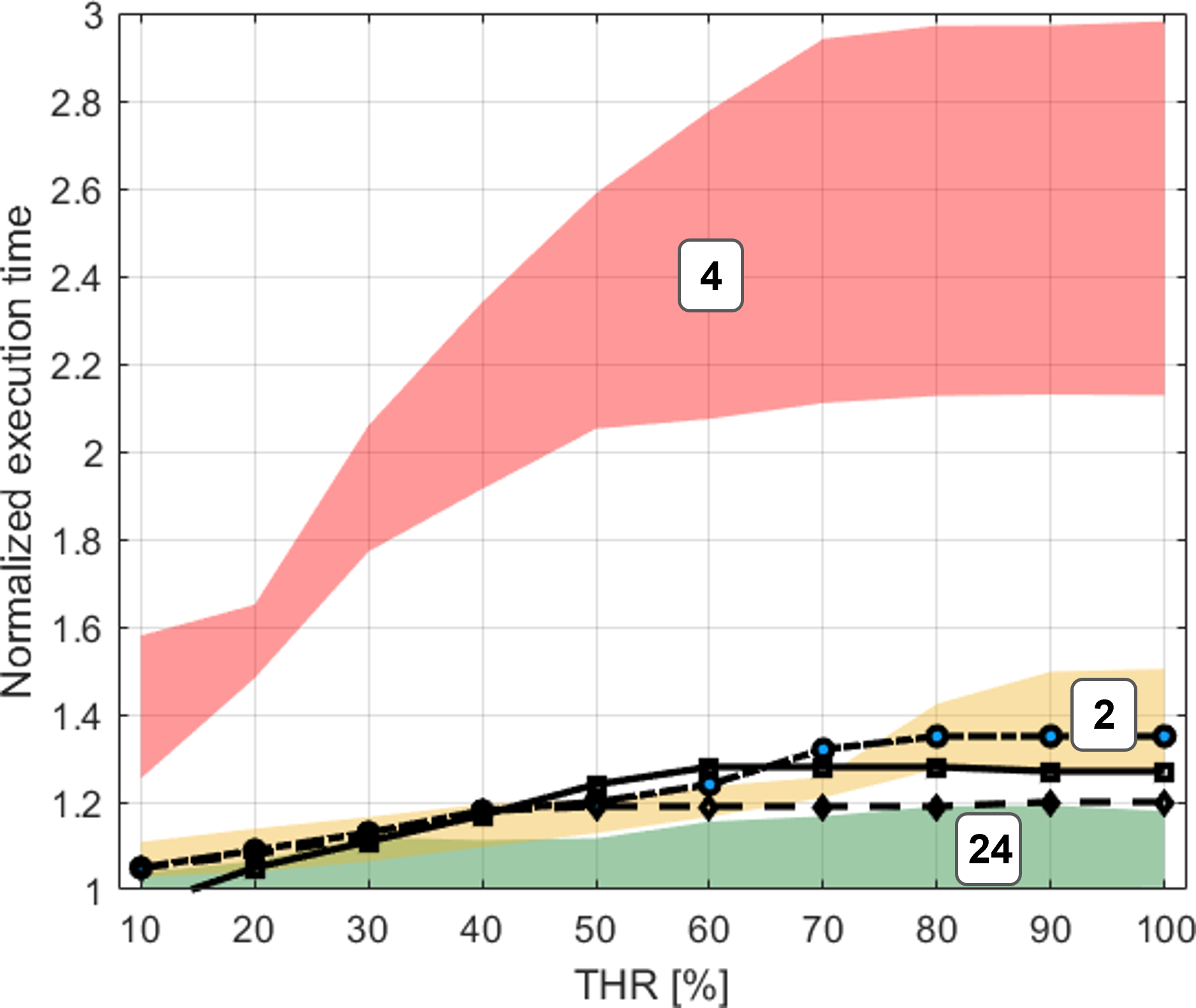}  
        \label{fig:NVIDIAa}
    \end{subfigure}
    \begin{subfigure}[c]{.28\textwidth}
        \centering
        % include first image
        \caption{\textbf{Interf=MEMSET}}
        \includegraphics[trim=20 0 0 0,clip,width=\columnwidth]{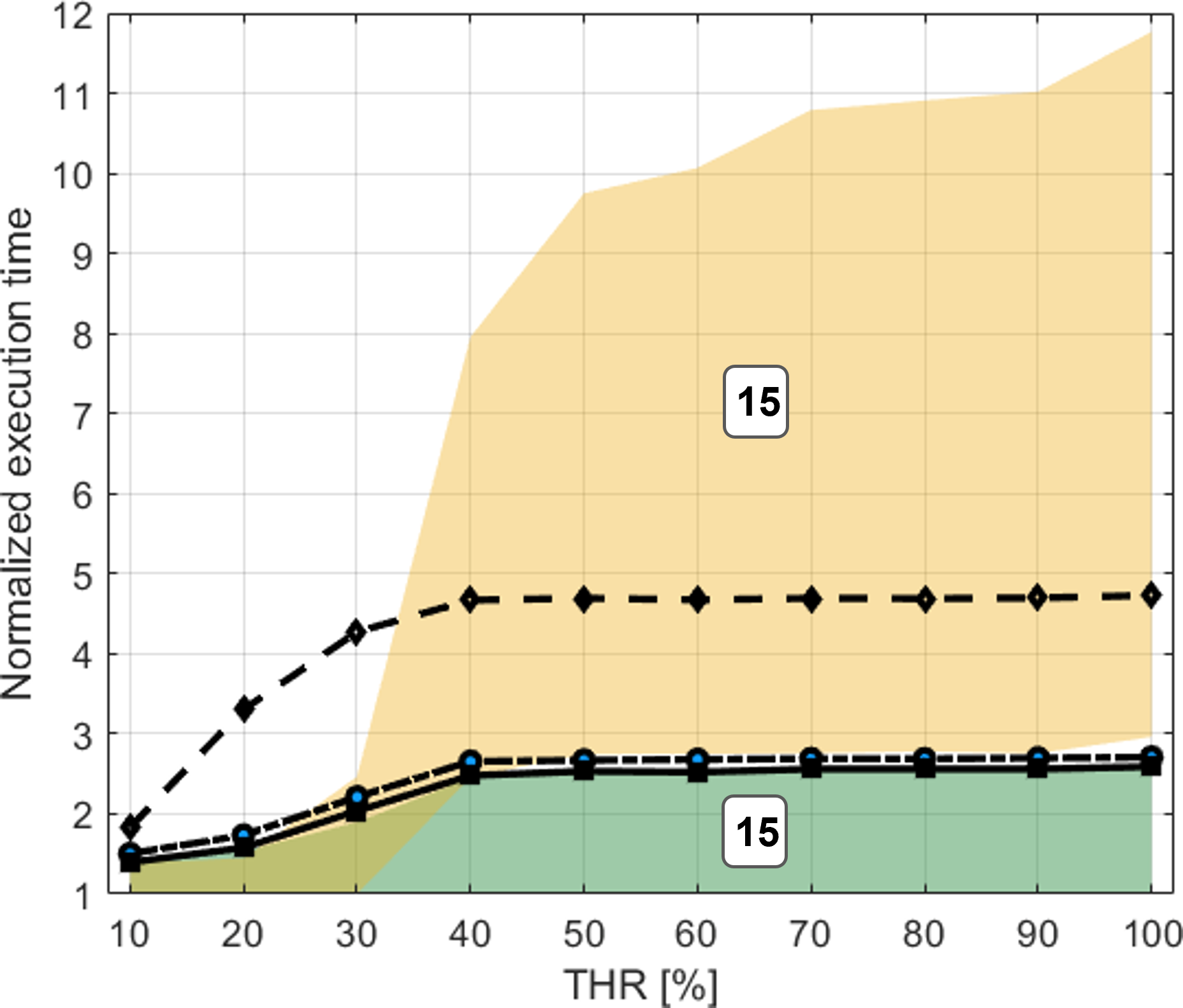}  
        \label{fig:NVIDIAb}
    \end{subfigure}
    \begin{subfigure}[c]{.28\textwidth}
        \centering
        % include first image
        \caption{\textbf{Interf=MEMCPY}}
        \includegraphics[trim=20 0 0 0,clip,width=\columnwidth]{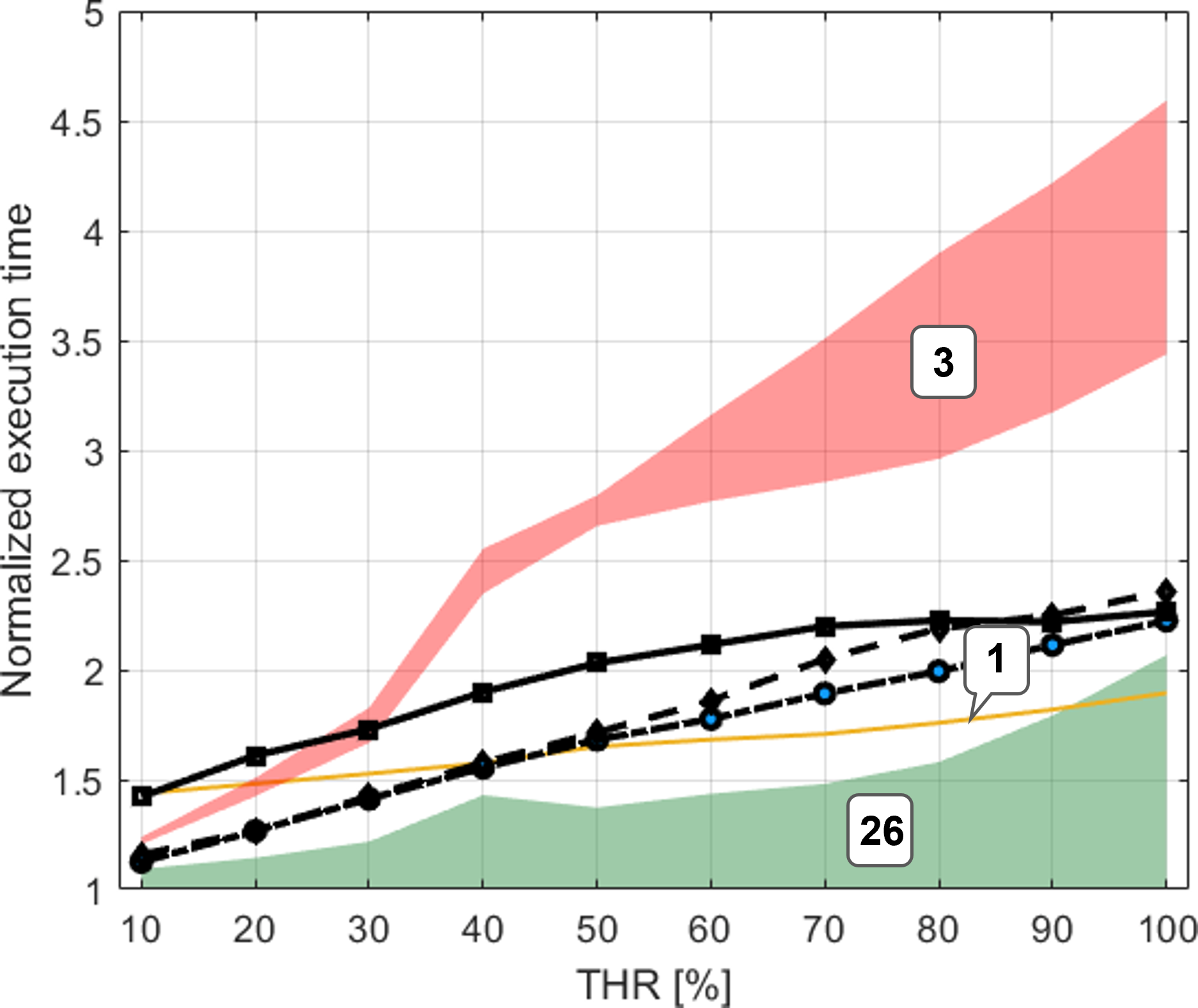}
        \label{fig:NVIDIAc}
    \end{subfigure}
    \begin{subfigure}[c]{.1\textwidth}
        \centering
        % include third image
        \includegraphics[width=\columnwidth]{imgs/Figure_2_legend.png}
        \label{fig:_tmp8}
    \end{subfigure}
    \caption{\textbf{Xilinx ZU9EG}. Execution time increase of the three synthetic benchmarks (curves) and the Polybench benchmarks (solored areas) running on the core under test with increasing interference from the other cores (THR\%). The workload executed by the interference cores is indicated above the plot.}
    \label{fig:interf_Xilinx}
\end{figure*}

Some previous works~\cite{10.1145/3548773,read_bcs,9203722} make the assumption that read-only synthetic benchmarks, which are programs executing memory operations at the highest speed possible, have to be the ones which either:
\begin{enumerate}
    \item Cause the highest amount of DRAM interference for other concurrently running tasks.
    \item Are the most slowed down by the effects of DRAM interference.
\end{enumerate}

Using such concepts without a proper study of the hardware is particularly problematic, since they could lead to a wrong characterization of the worst-case, which, in turn, could mean that certain conditions might cause slowdowns greater than expected for important time-sensitive applications.

\subsection{Contributions}
Our research analyzed the effects of different synthetic memory-intensive benchmarks running alongside other tasks (Polybench) on two separate HeSoCs: the FPGA-based Xilinx ZU9EG and the GPU-based NVIDIA TX2.
Our results prove that:
\begin{enumerate}
    \item The type of program causing the highest amount of interference is actually hardware-dependant.
    \item The type of program most slowed down by the effects of DRAM interference is not guaranteed to be read-only synthetic benchmarks.
\end{enumerate}

\section{Synthetic benchmarks}
Synthetic benchmarks are programs which can be fine-tuned to emulate different kinds of programs based on certain parameters.
For our tests on DRAM interference generation, we decided on using the following configurations:
\begin{itemize}
    \item \emph{READ\_MISS}: Loads only, at the maximum possible speed. What was believed to be the worst-case in Section \ref{sec:intro}. It produces a \emph{100 \% read} traffic.
    \item \emph{MEMSET}: A series of consecutive stores, executed at the maximum possible speed. It produces a \emph{100 \% write} traffic.
    \item \emph{MEMCPY}: Consecutive loads and stores. It produces a \emph{50\% read - 50 \% write} traffic.
\end{itemize}

These synthetic benchmarks were then set to run on three CPU cores on the two platform as \emph{Background Interference Generators}. On the remaining core, another task (either a Polybench, or one of the Synthetic Benchmarks) was set to run, and the amount of slowdown it experienced was measured.
\section{Experimental Evaluation}
We ran our experiments on both the NVIDIA TX2 (Fig. \ref{fig:interf_NVIDIA}) and the Xilinx ZU9EG (Fig. \ref{fig:interf_Xilinx}) for varying levels of interference intensity (\textbf{THR\%}). The \emph{Above} region (red) identifies Polybench which always experience a worse slowdown than READ\_MISS. The \emph{Crossing} region (yellow) is for the Polybench which are subject to a worse slowdown than READ\_MISS only at certain points. Finally, the \emph{Below} region (green) is for Polybench which are never more slowed down than READ\_MISS, which is what previous literature actually accounted for.
The results definitively prove that read only synthetic benchmarks (\emph{READ\_MISS}) are not:
\begin{enumerate}
    \item The ones which cause the highest amount of interference, as can be seen by \emph{MEMSET}'s highest slowdown being around 12x on the ZU9EG, compared to \emph{READ\_MISS}'s 3x. On the TX2, \emph{MEMCPY} causes the highest amount of interference, instead. It's important to note that there may be even worse interference patterns, that we have not yet found.
    \item The programs which are slowed down the most. Depending on the memory access pattern of the interference, different synthetic benchmarks become the worst case. Not only that, but there are certain Polybench which reach even higher degrees of slowdown for all the different interference causing traffic patterns. This, however, is due to Cache events, and not exclusively DRAM interference.
\end{enumerate}

For the ZCU102, the combined effect of these factors makes it so the real worst-case is actually subject to a 12x slowdown factor, instead of the 1.3x which could be observed when using \emph{READ\_MISS} as both the benchmark under test, and the program causing the interference.

% \begin{figure*}
%     \begin{subfigure}[c]{.3\textwidth}
%         \centering
%         % include second image
%         \caption{\textbf{Interf=READ\_MISS}}
%         \includegraphics[width=\columnwidth]{imgs/Figure_2_zcu102_regions.png}  
%         \label{fig:XILINXa}
%     \end{subfigure}
%     \begin{subfigure}[c]{.28\textwidth}
%         \centering
%         % include second image
%         \caption{\textbf{Interf=MEMSET}}
%         \includegraphics[trim=20 0 0 0,clip,width=\columnwidth]{imgs/Figure_3_zcu102_memset.png}  
%         \label{fig:XILINXb}
%     \end{subfigure}
%     \begin{subfigure}[c]{.28\textwidth}
%         \centering
%         % include second image
%         \caption{\textbf{Interf=MEMCPY}}
%         \includegraphics[trim=20 0 0 0,clip,width=\columnwidth]{imgs/Figure_5_memcpy_zcu102.png}  
%         \label{fig:XILINXc}
%     \end{subfigure}
%     \begin{subfigure}[c]{.1\textwidth}
%         \centering
%         % include third image
%         \includegraphics[width=\columnwidth]{imgs/Figure_2_legend.png}  
%         \label{fig:label_XILINX}
%     \end{subfigure}
%     \caption{\textbf{Xilinx ZU9EG}. Execution time increase of the three synthetic benchmarks (curves) and the Polybench benchmarks (solored areas) running on the core under test with increasing interference from the other cores (THR\%). The workload executed by the interference cores is indicated above the plot.}
%     \label{fig:interf_XILINX}
% \end{figure*}
%\input{sections/experiments}
\section{Conclusion}

\begin{figure}[]
    \includegraphics[width=0.5\textwidth]{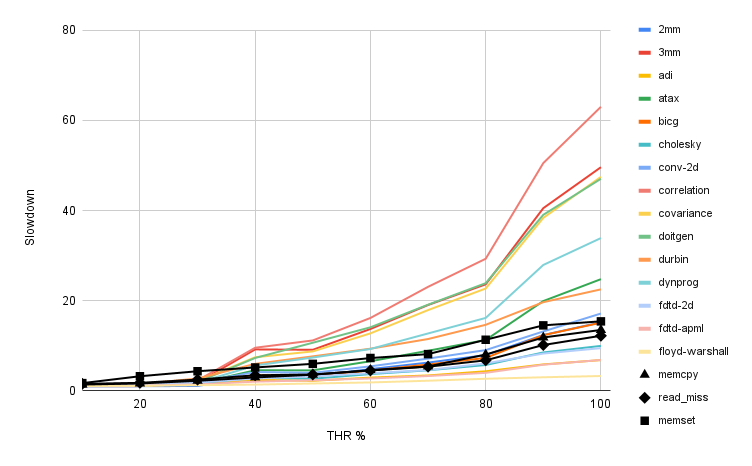}
        \caption{Slowdown for certain Polybench tasks on the Xilinx ZU9EG, with 3 other CPU cores running MEMSET, and 3 FPGA cores producing RW traffic}\label{fig:fpga-combo}
\end{figure}

Our research makes it clear that proper worst-case characterization is important when developing memory contention mitigation techniques. While being focused on DRAM interference is important, Cache events must be also accounted for, as they can cause regular tasks to be subject to more interference than the most memory intensive ones (Red/Yellow regions in Fig. \ref{fig:interf_NVIDIA} and \ref{fig:interf_Xilinx}). On the Xilinx ZU9EG (Fig. \ref{fig:interf_Xilinx}), the situation is particularly egregious, as the worst-case for our tests is actually more than 9$\times$ worse than \emph{READ\_MISS}. All of these evaluations have been done with the accelerators turned off, in order to create a valid comparison of the effects of bad worst-case characterization on different HeSoCs. However, we have observed that programs can experience higher degrees of slowdown when the platform-specific hardware is used. For example, on the Xilinx ZU9EG, when the FPGA cores are used to generate traffic, the total slowdown a program can experience can be greater than 60$\times$ (Fig. \ref{fig:fpga-combo}) if the CPU cores are also executing \emph{MEMSET}. Any proposal which fails to account for these kind of scenarios cannot realistically state to have covered the worst-case slowdown which a program can experience on these kinds of platforms.

\bibliographystyle{IEEEtran}
\bibliography{bibliography}
\end{document}